\documentclass[
showpacs,preprintnumbers,amsmath,amssymb]{revtex4}

\begin{document}
\title{Nonergodic solutions of the generalized Langevin equation}
\author{A.V. Plyukhin}
\email{aplyukhin@anselm.edu}
 \affiliation{ Department of Mathematics,
Saint Anselm College, Manchester, New Hampshire 03102, USA 
}

\date{\today}

\begin{abstract}
It is known that in the regime of superlinear diffusion,
characterized by zero integral friction (vanishing integral of the memory 
function), 
the generalized Langevin equation may 
have non-ergodic solutions that do not relax to equilibrium values. 
It is shown that  the equation may have non-ergodic (non-stationary) solutions  
even if the integral of the  memory function is finite and 
diffusion is normal.

\end{abstract}


\maketitle
There is hardly anything more important to say about a statistical 
mechanical system than whether it is ergodic or not.
In general the question is notoriously difficult,  
yet   for certain  classes of stochastic systems the criteria of ergodicity breaking may be 
remarkably simple~\cite{Barkai1,Barkai2,Costa,Lapas,Bao}. 
This is so, or so it would appear,
 for
stochastic dynamics described by the generalized Langevin equation (GLE)
\begin{eqnarray}
\frac {d A(t)}{dt}=-\int_0^t d\tau \, M(t-\tau)\,A(\tau) + F(t),
\label{GLE}
\end{eqnarray}
which governs
a dynamical variable $A$ of a classical system coupled to a  thermal bath with many
degrees of freedom in the absence of external forces~\cite{Zwanzig}.  
The ``random'' force $F(t)$ is zero centered $\langle F(t)\rangle=0$, 
not correlated with the initial value of  $A$
\begin{eqnarray}
\langle A(0)F(t)\rangle =0,
\label{aux1}
\end{eqnarray}
and related with the dissipative memory function $M(t)$ 
through
the fluctuation-dissipation theorem
\begin{eqnarray}
\langle F(0)F(t)\rangle=\langle A^2\rangle\,M(t). 
\label{FDT}
\end{eqnarray}
We shall also assume  the asymptotic 
vanishing of correlations of the random force 
\begin{eqnarray}
\lim_{t\to\infty}\langle F(0)F(t)\rangle=\lim_{t\to\infty}M(t)=0, 
\label{cond_0}
\end{eqnarray}
which is typical for  irreversible stochastic processes.
It appears to be a common belief that, given conditions (\ref{FDT}) and (\ref{cond_0}),
solutions of the GLE (\ref{GLE}) describe  ergodic 
relaxation to thermal equilibrium, unless  
the Laplace transform of the memory function 
 $\tilde M(s)=\int_0^\infty dt\,e^{-st}M(t)$ 
has a specific asymptotic behavior.
Namely, it was shown in~\cite{Costa,Lapas,Bao} that    
the condition of ergodicity breaking for GLE systems has the form
\begin{eqnarray} 
\tilde M(s)\sim s^\delta,\,\,\,\,\delta \ge 1, 
\quad\quad
\mbox{as}
\quad
s\to 0.
\label{condition1}
\end{eqnarray}  
This condition  implies 
the vanishing integral of the memory function
\begin{eqnarray}
\int_0^\infty dt\,M(t)= \tilde M(0)=0.
\label{condition}
\end{eqnarray}
If the targeted variable $A$ is the velocity of a Brownian particle, 
condition  (\ref{condition}) 
corresponds to  anomalous
diffusion 
when the mean-square displacement $\langle x^2(t)\rangle$
of the particle increases with time as 
$t^\alpha$ with $\alpha>1$ (superdiffusion)~\cite{Morgado,Hanggi}. 
The relation (\ref{condition}) is not very common, but not unrealistic. 
For instance, it was found 
to hold  for a particle
interacting with longitudinal phonons in liquids in the limit of 
zero temperature~\cite{Frenkel}.
It should perhaps be noted  that while the condition of 
ergodicity breaking (\ref{condition1})
invariably implies superdiffusion, the converse is not true: for 
$\tilde M(s)\sim s^\delta$ with $0<\delta<1$ the condition of 
superdiffusion (\ref{condition}) is satisfied yet
solutions of the GLE (\ref{GLE}) are ergodic (\cite{Costa}, 
see also  Eq.(\ref{cond_laplace}) below).

The purpose of this paper is to show that
the condition of ergodicity breaking
in the form (\ref{condition1}) is too restrictive.
It will be demonstrated that the GLE may have non-ergodic solutions even if
the  memory function does not follow the asymptotic form 
(\ref{condition1}),  $\tilde M(0)=\int_0^\infty M(t)\, dt$ 
is finite, and diffusion is normal.

We begin by briefly recapitulating the derivation
of condition (\ref{condition1}) which may differ depending 
on a type of averaging 
$\langle ...\rangle$
in relations (\ref{aux1}) and (\ref{FDT}).
When the GLE is 
derived with the Mori's projection operator technique~\cite{Zwanzig},
the system is usually assumed to be in thermal equilibrium  with the bath, 
and the averaging 
in equations (\ref{aux1}) and (\ref{FDT}) is over the ensemble of 
initial conditions for the composition of  
the system and the bath in mutual thermal equilibrium. 
In this case  it is natural to use the GLE to 
evaluate 
the equilibrium correlation function $\langle A(0)A(t)\rangle$.
Its normalized   form
\begin{eqnarray}
C(t)=\frac{\langle A(0)A(t)\rangle}{\langle A^2(0)\rangle}
\end{eqnarray} 
satisfies the equation
\begin{eqnarray}
\frac {d C(t)}{dt}=-\int_0^t d\tau \, M(t-\tau)\,C(\tau)
\label{Ceq}
\end{eqnarray}   
with  the initial condition $C(0)=1$, and has a
Laplace transform
\begin{eqnarray}
\tilde C(s)=\frac{1}{s+\tilde M(s)}.
\label{C}
\end{eqnarray}
The connection to ergodic properties is given 
by  Khinchin's theorem~\cite{Khinchin} (see also~\cite{Barkai2}),
which states that the stationary  process $A(t)$ is ergodic if
the correlation function factorises and, for a  zero-centered process,
vanishes in the long time limit
\begin{eqnarray}
\lim_{t\to\infty}C(t)=\frac{\langle A(0)\rangle\langle A(t)\rangle}
{\langle A^2(0)\rangle}=0.
\label{KT}
\end{eqnarray}
Although in  Mori's GLE the random force $F(t)$, and therefore $A(t)$,   
are not necessarily stationary and zero centered, 
 we shall assume that these properties do hold. 
Then  equations (\ref{C}), (\ref{KT}) and the limit value theorem
\begin{eqnarray}
\lim_{t\to\infty} C(t)=\lim_{s\to 0}s\,\tilde C(s)
\label{LMT}
\end{eqnarray}
give the condition of ergodicity breaking  in the following  form
\begin{eqnarray}
\lim_{s\to
  0}\frac{s}{s+\tilde M(s)}\ne 0,
\label{cond_laplace}
\end{eqnarray} 
which leads to  condition (\ref{condition1}).

A slightly different approach is to apply for 
a particular and very popular class of models  where the
random force $F(t)$ does not depend on $A$.
This is the case, for instance, when $A$ is the momentum of 
a Brownian particle which is  
bilinearly coupled to the bath comprised of harmonic 
oscillators~\cite{Zwanzig}. 
For this problem, often referred to as the Caldeira-Leggett model,  relation (\ref{aux1}) is satisfied trivially,
the random force is stationary (for the infinite bath), and 
the fluctuation dissipation theorem
takes the form 
\begin{eqnarray}
\langle F(0)F(t)\rangle_0=\langle A^2\rangle\,M(t), 
\label{FDT2}
\end{eqnarray}
where the 
the average $\langle ...\rangle_0$ 
is taken over bath variables only. 
The latter allows one to use the GLE to evaluate not only the equilibrium correlation function, but
also the second moment $\langle A^2(t)\rangle_0 $ which characterizes the process of thermalization
of the system which  at the moment $t=0$ is put in contact with the equilibrium thermal bath.
Compared to  Mori's approach, this  is a more general problem since the initial equilibrium of the system and the bath is not assumed.
One can  show  that  the system does not thermalize,
\begin{eqnarray}
\lim_{t\to\infty}\langle A^2(t)\rangle_0\ne\langle A^2\rangle,
\label{th}
\end{eqnarray}
under the same condition as 
that for ergodicity breaking discussed above.
Indeed, using Laplace transformation the solution of the GLE (\ref{GLE}) can be written in the form 
\begin{eqnarray}
A(t)=A(0)\,C(t)+\int_0^t d\tau \,C(t-\tau) \,F(\tau).
\label{solution1}
\end{eqnarray}
where  the response function $C(t)$ has the  transform given by Eq.(\ref{C}) 
and therefore coincides with the correlation function 
for  Mori's GLE and  satisfies Eq.(\ref{Ceq}).
By squaring and averaging  solution (\ref{solution1}) over bath variables, 
and also using stationarity of $F(t)$ and the fluctuation-dissipation relation (\ref{FDT2})
\begin{eqnarray}
\langle F(t_1)F(t_2)\rangle_0=\langle A^2\rangle\,M(|t_1-t_2|),
\end{eqnarray}
one gets
\begin{eqnarray}
\langle A^2(t)\rangle_0 &=& A^2(0)\,C^2(t) 
+2\,\langle A^2\rangle\int_0^t \!d\tau_1 C(\tau_1)\!\int_0^{\tau_1} \!d\tau_2\,
C(\tau_2) M(\tau_1\!-\!\tau_2). 
\end{eqnarray}
Using (\ref{Ceq}), this equation can be written as  
\begin{eqnarray}
\!\langle A^2(t)\rangle_0 = A^2(0)\,C^2(t)-
2\,\langle A^2\rangle\,\,\int_0^t d\tau_1 C(\tau_1)\,\dot C(\tau_1),
\nonumber
\end{eqnarray}
and eventually one obtains~\cite{Masoliver}
\begin{eqnarray}
\langle A^2(t)\rangle_0 &=& A^2(0)\,C^2(t)+\langle A^2\rangle\,[1-C^2(t)].
\label{basic}
\end{eqnarray}
The system does not thermalize if the response function
does not vanish in the long time limit, 
$\lim_{t\to\infty} C(t)\ne 0$, 
which again gives the conditions  (\ref{cond_laplace}) and (\ref{condition1}).

The above reasoning  was based on the 
limit value theorem (\ref{LMT}) 
which is only valid if the system reaches a stationary state and the long time limit for $C(t)$ does exist.
One might suggest that this is always the case provided the random force is irreversible in the sense
that the correlation function  $\langle F(0)F(t)\rangle$ and  the memory kernel $M(t)$  vanish as $t\to\infty$.  
Let us show that this assumption is incorrect: It is possible to construct memory functions  $M(t)$  which vanish at long times, but 
the corresponding functions $C(t)$, related to $M(t)$  
by Eq. (\ref{Ceq}) or (\ref{C}),  do not have a long time limit. 
The condition of superdiffusion $\int_0^\infty M(t)\,dt=0$ is not required.

As an example, let us 
consider a class of memory functions
with the Laplace transform 
\begin{eqnarray}
\tilde M(s)=-s+\frac{s^2+\omega^2}{f(s)},
\label{guess}
\end{eqnarray} 
where $\omega$ is real, and $f(s)$ is an analytic function at $s=\pm i\omega$.
The corresponding transform for the correlation or
response function $C(t)$, given by (\ref{C}),  is
\begin{eqnarray}
\tilde C (s)=\frac{f(s)}{s^2+\omega^2}.
\end{eqnarray} 
It  has simple poles at $s=\pm i\omega$ on the imaginary axis and therefore
the original $C(t)$ contains terms oscillating with frequency $\omega$ 
and does not  reach 
a stationary value  as $t\to\infty$. 
It is not immediately obvious, however,  whether it is possible 
to construct a  function $f(s)$ which ensures that the memory kernel 
behaves in a physically  reasonable way. 
There are several conditions to satisfy. First,
as a Laplace transform must vanish in the limit
$s\to\infty$, we must require, in view of (\ref{guess}), that  
\begin{eqnarray}
f(s)\sim s, \qquad \mbox{as $s\to\infty$}.
\label{cond1}
\end{eqnarray}
The second condition is the 
asymptotic vanishing of correlations (\ref{cond_0}), 
\begin{eqnarray}
\lim_{t\to\infty}M(t)=\lim_{s\to 0}s\,\tilde M(s)=0,
\label{cond_v}
\end{eqnarray}
which leads to the asymptotic constraint
\begin{eqnarray}
f(s)\sim s^r, \qquad r<1\qquad \mbox{as $s\to 0$}.
\label{cond_vv}
\end{eqnarray}
The third condition, which is more difficult to handle than the other two, 
is that 
the memory function must not exceed its initial value, 
\begin{eqnarray}
|M(t)|\le M(0), \qquad \mbox{for $ t>0$}.
\label{cond_2}
\end{eqnarray}
This is because $M(t)$ is essentially the correlation function
of the stationary stochastic process $F(t)$,  which satisfies the inequality 
\begin{eqnarray}
\langle [F(0)-F(t)]^2\rangle=
2\langle F^2(0)\rangle-2\langle F(0)F(t)\rangle\ge 0.
\nonumber
\end{eqnarray}
Condition (\ref{cond_2}) cannot be formulated as that for the Laplace
transform $\tilde M(s)$, which makes the choice of $f(s)$ 
in Eq.(\ref{guess})  not quite straightforward. It turns out that 
the simplest function $f(s)$ that can be made consistent with 
all three conditions (\ref{cond1}), (\ref{cond_vv}), and 
(\ref{cond_2}) is
\begin{eqnarray}
f(s)=\frac{s^2+as+b}{s+c},
\label{class}
\end{eqnarray}
with certain restrictions on  the real constants $a,b,$ and $c$. 
In this  case 
\begin{eqnarray}
\tilde C(s)=\frac{s^2+as+b}{(s+c)(s^2+\omega^2)},
\label{C_L}
\end{eqnarray}
while the transform of the the memory function  (\ref{guess}) 
can be written  in 
the form
\begin{eqnarray}
\tilde M(s)=\alpha+\frac{\beta s+\gamma}{s^2+as+b},
\label{M_L}
\end{eqnarray}
where 
\begin{eqnarray}
\alpha=c-a,\qquad
\beta=\omega^2-b-a(c-a),\qquad
\gamma=\omega^2c-b(c-a).
\label{aux4}
\end{eqnarray}
The inverse Laplace transformation $\mathcal L^{-1}$ of (\ref{M_L}) gives the memory function 
as a sum of the Dirac delta-function and a non-singular part,  
\begin{eqnarray}
M(t)=\alpha\,\delta(t)+m(t).
\label{M}
\end{eqnarray}
Conditions (\ref{cond_v}) and (\ref{cond_2}) are satisfied 
if the singular part is positive ($\alpha>0$) and the non-singular function
\begin{eqnarray}
m(t)=\mathcal L^{-1}\left\{
\frac{\beta s+\gamma}{s^2+as+b}
\right\} 
\end{eqnarray}
vanishes as $t\to\infty$.  This sets  the constraints
\begin{eqnarray}
c>a>0,\qquad b\ne 0,
\label{aux5}
\end{eqnarray}
while $\omega$ is still an arbitrary parameter.

As an illustration consider 
the set of parameters $a=b=1$, $c=2$, and $\omega^2=2$.
Then  equations (\ref{aux4}) give  $\alpha=1$, $\beta=0$,  $\gamma=3$, and 
 transforms (\ref{C_L}) and (\ref{M_L}) read  
\begin{eqnarray}
\tilde M(s)=1+\frac{3}{s^2+s+1},\qquad
\tilde C(s)=\frac{s^2+s+1}{(s^2+2)(s+2)}.
\end{eqnarray}
Respectively, in this case the memory function is 
\begin{eqnarray}
M(t)=\delta(t)+2\sqrt{3}\,e^{-t/2}\,\sin\left(\frac{\sqrt{3}}{2}\,t\right),
\end{eqnarray}
and satisfies both conditions (\ref{cond_v}) and (\ref{cond_2}), 
while the correlation or response function 
\begin{eqnarray}
C(t)&=&\frac{1}{2}\,e^{-2t}+\frac{1}{2}\,\cos\left(\sqrt{2}\, t\right)
\end{eqnarray}
does not reach a long time limit. Observe that the non-singular part of the memory function $m(t)$  
increases at $t=0$.  As one can check, 
for the given class of memory functions (\ref{guess}), 
and under  restriction  (\ref{cond_0}),  
the  property  $m'(0)>0$ is generic.
Therefore  the  presence of a singular term in  $M(t)$ is essential: 
if the delta function is absent in (\ref{M}) ($\alpha=0$), then 
the condition (\ref{cond_2}) cannot be met. 

Needless to say, 
the condition of superdiffusion  
$\int_0^\infty dt\,M(t)=\tilde M(0)=0$, is not implied
in our construction. It follows from (\ref{M_L}) that
$\tilde M(0)=\omega^2c/b$,  and so, unless $\omega=0$,
$\tilde M(0)$ is finite and  diffusion is normal. 
Since $b\ne 0$ due to (\ref{aux5}), 
the regime of  subdiffusion $\lim_{s\to 0}\tilde M(s)=\infty$~\cite{Morgado}
does not occur for   memory functions  of type (\ref{M_L}).

Summarizing, it is shown that stochastic dynamics governed by the
generalized Langevin equation 
may be non-dissipative in the regime of normal diffusion, and
thus superdiffusion is not a necessary condition for 
ergodicity breaking, as often assumed in literature.
In our showcase example the memory function     
consists  of a delta-peak and a long, generally nonmonotonic tail.
A possibility  of non-ergodic dynamics
generated by a  
noise with physically more realistic 
non-singular correlations   remains to be examined.
In this case the time-reversal symmetry requires that
the exact memory and autocorrelation functions  must be  even in  time~\cite{ad}. 
Neither this additional constraint, nor the condition of subdiffusion
(divergence of $\tilde M(s)$ as $s\to 0$) can be met by the simple 
class of memory functions considered in this paper.

I thank Gregory Buck and Stephen Shea for discussion  and the anonymous referees 
for important corrections.



\begin{thebibliography}{99}

\bibitem{Barkai1} A. Rebenshtok and E. Barkai, Phys. Rev. Lett. {\bf 99},
 210601 (2007);
S. Burov and E. Barkai, {\it ibid.} {\bf 98}, 250601 (2007); 
W. Deng and E. Barkai, Phys. Rev. E 79, 011112 (2009).

\bibitem{Barkai2}
S. Burov, R. Metzler, and E. Barkai, PNAS {\bf 107}, 13228 (2010).



\bibitem{Costa} I. V. L. Costa, R. Morgado, M. V. B. T. Lima, and 
F. A. Oliveira, Europhys. Lett. {\bf 63}, 173 (2003).


\bibitem{Bao} J. D. Bao, P. Hanggi, and Y. Z. Zhuo, 
Phys. Rev. E {\bf 72}, 061107 (2005).


\bibitem{Lapas} L. C. Lapas, R. Morgado, M. H. Vainstein, J. M. Rubi, and
  F. A. Oliveira, Phys. Rev. Lett. {\bf 101}, 230602 (2008). 


\bibitem{Zwanzig} R. Zwanzig, {\it Nonequilibrium statistical mechancis}, 
Oxford University Press, New York (2001).

\bibitem{Khinchin} A. I. Khinchin, {\it Mathematical Foundations of
  Statistical Mechanics}, Dover, New York (1949).

\bibitem{Morgado} R. Morgado, F. A. Oliveira, G. G. Batrouni, and A. Hansen,
Phys. Rev. Lett. {\bf 89}, 100601 (2002).


\bibitem{Hanggi} P. Siegle, I. Goychuk, P. Talkner, and P. Hanggi,
  Phys. Rev. E {\bf 81}, 011136 (2010); P. Siegle, I. Goychuk, P. Hanggi, Phys. Rev. Lett. {\bf 105}, 100602 (2010).

\bibitem{Frenkel} G. Frenkel and M. Schwartz, Europhys. Lett. {\bf 50}, 
628 (2000).

 
\bibitem{Masoliver} J. M. Porr{\`a}, K.-G. Wang, and J. Masoliver,
  Phys. Rev. E {\bf 53}, 5872 (1996). 

\bibitem{ad}
M. H. Lee, Phys. Rev. Lett. {\bf 51} 1227 (1983);
M. H. Vainstein, I. V. L. Costa, R. Morgado, and F. A. Oliveira, 
Europhys. Lett. 73, 726 (2006).







\end{thebibliography}
\end{document}